\documentclass[conference]{IEEEtran}
\IEEEoverridecommandlockouts
\usepackage{cite}
\usepackage{amsmath,amssymb,amsfonts}
\usepackage{algorithmic}
\usepackage{algorithm}
\usepackage{graphicx}
\usepackage{textcomp}
\usepackage{xcolor}
\def\BibTeX{{\rm B\kern-.05em{\sc i\kern-.025em b}\kern-.08em
    T\kern-.1667em\lower.7ex\hbox{E}\kern-.125emX}}
\begin{document}

\title{Joint Semantic Coding and Routing for Multi-Hop Semantic Transmission in LEO Satellite Networks\\
\thanks{This work is supported by the National Natural Science Foundation of China(No.~U25B2033, No.~U23A20275 and No.~62171072 ).}
}

\author{\IEEEauthorblockN{Hong Zeng}
\IEEEauthorblockA{\textit{School of Communications and} \\
\textit{Information Engineering} \\
\textit{Chongqing University of Posts and}\\
\textit{Telecommunications} \\
Chongqing, China \\
s240101201@stu.cqupt.edu.cn}
\and
\IEEEauthorblockN{Jiangtao Luo\textsuperscript{*}}
\IEEEauthorblockA{\textit{School of Communications and} \\
\textit{Information Engineering} \\
\textit{Chongqing University of Posts and}\\
\textit{Telecommunications} \\
Chongqing, China \\
luojt@cqupt.edu.cn}
\and
\IEEEauthorblockN{Yongyi Ran\textsuperscript{*}}
\IEEEauthorblockA{\textit{School of Communications and} \\
\textit{Information Engineering} \\
\textit{Chongqing University of Posts and}\\
\textit{Telecommunications} \\
Chongqing, China \\
ranyy@cqupt.edu.cn}
\thanks{\textsuperscript{*} Jiangtao Luo and Yongyi Ran are the Corresponding authors (Email: luojt@cqupt.edu.cn, ranyy@cqupt.edu.cn).}
}

\maketitle

\begin{abstract}
Low Earth Orbit satellite networks pose significant challenges to multi-hop semantic transmission because rapidly changing topology, link variability, and queue dynamics make end-to-end performance jointly depend on routing, relay processing, and semantic payload adaptation. Existing studies usually optimize routing or semantic transmission separately and are therefore not well suited to dynamic satellite scenarios under local observations. To address this issue, this paper proposes GraphJSCR, a graph-based joint routing and semantic coding method for multi-hop semantic transmission in dynamic Low Earth Orbit satellite networks. The satellite constellation is modeled as a time-varying directed graph, and the forwarding process is formulated as a partially observable sequential decision problem. A graph representation learning module is designed to encode local topology, link status, queue conditions, packet context, and semantic transmission states. Based on the learned representation, the proposed decision network jointly determines next-hop selection, relay processing level, and semantic transmission budget to balance end-to-end semantic quality and transmission delay. The semantic encoder-decoder is developed with reference to the SwinJSCC framework. Simulation results demonstrate that GraphJSCR achieves faster convergence and a better tradeoff between semantic fidelity and transmission efficiency than benchmark methods.
\end{abstract}


\section{Introduction}

Low Earth Orbit (LEO) satellite networks are expected to play a key role in future integrated space-air-ground communication systems because of their wide-area coverage, flexible deployment, and low dependence on terrestrial infrastructure\cite{b1}. However, their high orbital mobility, rapidly time-varying topology, and frequent inter-satellite link switching make reliable multi-hop transmission highly challenging. Meanwhile, emerging satellite services increasingly involve image- and video-oriented traffic, for which conventional bit-level reliable delivery is often inefficient in terms of transmission overhead, latency, and task effectiveness\cite{b2}. This motivates semantic-oriented transmission, where the objective is to preserve task-relevant information rather than exact bit recovery.

Although recent advances in intelligent routing for LEO networks\cite{b3} and semantic communication based on Deep JSCC\cite{b2} and its Swin Transformer variant\cite{b4} provide useful foundations, they do not directly resolve the core challenge of multi-hop semantic transmission in dynamic LEO networks. In a multi-hop semantic system, end-to-end performance is jointly affected by path selection, relay behavior, and semantic transmission budget adaptation. Forward-only relaying can cause representation mismatch and accumulation of distortions across heterogeneous links, whereas decode-and-reencode relaying can improve robustness at the cost of additional delay and resource consumption onboard \cite{b5}. Therefore, the essential problem is not routing alone or coding alone, but how to coordinate forwarding and semantic processing over multiple hops under dynamic topology and local observations.

Existing studies have started to uncover the coupling in multi-hop semantic transmission. Early efforts mainly focused on characterizing semantic degradation, whereas more recent works have moved toward active relay design. For example, \cite{b7} introduced residual compensation to reduce distortion accumulation in multi-hop image transmission. On the relay side, \cite{b8} leveraged semantic-state maintenance for efficient forwarding, \cite{b9} enhanced robustness via predictive feature extraction, and \cite{b10} extended semantic relaying to multi-source scenarios through relay-side fusion of critical semantic information. Cross-layer design has also been investigated in recent years, \cite{b11} revealed the coupling between semantic encoding and routing constraints, \cite{b13} proposed a task-oriented semantic delivery framework based on local topological information for heterogeneous satellite networks, and \cite{b14} employed hybrid DeepJSCC with adaptive compression-and-forward operations to combat performance degradation over unstable links. While these studies provide valuable insights, the problem of developing a unified distributed framework for dynamic LEO satellite networks remains relatively underexplored, especially under local observations where forwarding decisions need to jointly consider routing and semantic coding.

To address this issue, this paper proposes GraphJSCR, a graph-based joint routing and semantic coding algorithm for multi-hop semantic transmission in dynamic LEO satellite networks. The main contributions can be summarized as follows: First, we formulate multi-hop semantic transmission in LEO satellite networks as a unified joint decision problem integrating routing, relay processing, and semantic coding control under local observations. Second, we design a graph-representation-based distributed decision framework that exploits local topological and transmission states for adaptive hop-by-hop optimization. Third, we build an evaluation mechanism and demonstrate the effectiveness of the proposed method in jointly improving semantic quality and transmission efficiency.

\begin{figure*}[t]
    \centering
    \includegraphics[width=0.75\textwidth]{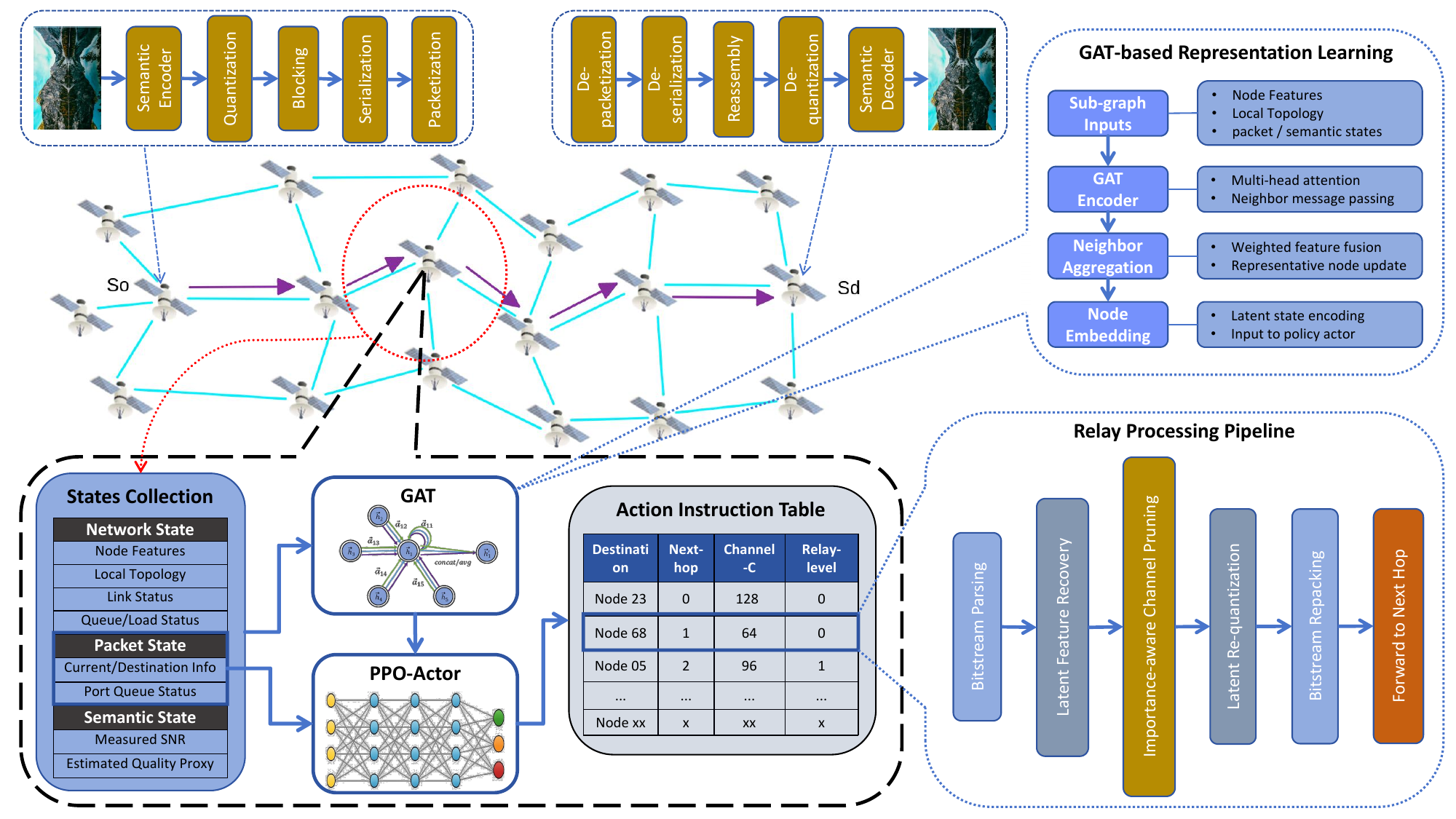}
    \caption{Overall framework of the proposed GraphJSCR method.}
    \label{fig:framework}
\end{figure*}

\section{Problem Formulation}

\subsection{System Architecture}
As illustrated in Fig.~\ref{fig:framework}, we consider multi-hop semantic image transmission over a dynamic LEO satellite network. At time slot $t$, the constellation is modeled as a time-varying directed graph
\begin{equation}
\mathcal{G}(t)=\big(\mathcal{V},\mathcal{E}(t)\big),
\end{equation}
where $\mathcal{V}$ denotes the set of satellites and $\mathcal{E}(t)$ denotes the set of available inter-satellite links (ISLs). Let $N=|\mathcal{V}|$ be the number of satellites. A semantic transmission task is initiated at a source satellite $s\in\mathcal{V}$ and terminated at a destination satellite $d\in\mathcal{V}$.

At the source node, the input image is first transformed into semantic features by the semantic encoder and then organized into packetized semantic payloads for multi-hop transmission over the dynamic LEO network. After reception at the destination, the semantic payload is restored and fed into the semantic decoder to reconstruct the semantic content. In our work, the semantic encoder-decoder is developed based on the SwinJSCC \cite{b4} framework, namely \textit{SwinJSCC w/ SA\&RA}.

Compared with conventional bit-level forwarding, the considered system additionally allows relay-side semantic processing at relay satellites. Therefore, the end-to-end transmission performance is jointly determined by the network path, relay processing behavior, and semantic payload budget along the multi-hop route.

\subsection{Relay Processing and Semantic Budget Adaptation}
Let $\mathbf{z}_k$ denote the encoded semantic latent representation of task $k$. Before transmission, $\mathbf{z}_k$ is packetized into semantic payload units and injected into the LEO network. For each relay satellite, the outgoing semantic payload can either be directly forwarded or further processed before transmission to the next hop.

We use $m_i$ to denote the relay processing mode/level at relay node $i$, and use $c_i$ to denote the semantic transmission budget allocated to the outgoing hop. The transmission budget can be interpreted as the effective semantic channel budget (e.g., Channel-$C$ in the implementation), which controls the amount of semantic information carried by the outgoing payload. Different relay levels correspond to different processing intensities on the semantic representation.

When relay processing is activated, the incoming semantic payload is processed through a relay pipeline including bitstream parsing, latent feature recovery, importance-aware channel pruning, latent re-quantization, and bitstream repacking. Denote the relay operator at node $i$ by
\begin{equation}
\tilde{\mathbf{z}}_{k,i}=\mathcal{R}_{m_i,c_i}\!\left(\mathbf{z}_{k,i}\right),
\end{equation}
where $\mathbf{z}_{k,i}$ and $\tilde{\mathbf{z}}_{k,i}$ are the incoming and outgoing semantic representations, respectively. The operator $\mathcal{R}_{m_i,c_i}(\cdot)$ adapts the semantic payload to downstream link and congestion conditions by jointly controlling relay-side processing and semantic budget allocation.

\subsection{Queue Model}
Each satellite maintains one receiving queue and multiple sending queues associated with its available outgoing directions. Newly arrived packets are first cached in the receiving queue and then dispatched to the corresponding sending queue according to the forwarding decision. Let $q_{i,p}(t)$ denote the length of the sending queue associated with port $p$ on satellite $i$ at time slot $t$. The queue evolution is modeled as
\begin{equation}
q_{i,p}(t+1)
=
\min\!\Big\{
\big[q_{i,p}(t)-o_{i,p}(t)\big]^+ + z_{i,p}(t),\;
q_{\max}
\Big\},
\end{equation}
where $z_{i,p}(t)$ and $o_{i,p}(t)$ denote the number of packets entering and leaving queue $(i,p)$ during time slot $t$, respectively, $[\cdot]^+=\max(\cdot,0)$, and $q_{\max}$ is the maximum queue capacity.

\subsection{Delay Model}
For a packet transmitted from satellite $i$ to satellite $j$ at time slot $t$, the propagation delay is given by
\begin{equation}
\tau_{i,j}^{\mathrm{prop}}(t)=d_{i,j}(t)/c_0,
\end{equation}
where $d_{i,j}(t)$ is the Euclidean distance between the two satellites and $c_0$ is the speed of light.

Let $R_{i,j}(t)$ denote the achievable transmission rate of link $(i,j)$ at time slot $t$, and let $b_{k,i}^{\mathrm{out}}(t)$ denote the outgoing payload size of task $k$ after possible relay-side semantic processing and budget adaptation at node $i$. Then, the transmission delay is
\begin{equation}
\tau_{k,i,j}^{\mathrm{tx}}(t)=b_{k,i}^{\mathrm{out}}(t)/R_{i,j}(t),
\end{equation}

The queuing delay of task $k$ at the selected sending queue $(i,p)$ is denoted by $\tau_{k,i,p}^{\mathrm{q}}(t)$, which depends on the instantaneous queue occupancy and service condition of the corresponding outgoing port. In addition, if relay-side semantic processing is enabled at node $i$, an extra processing delay $\tau_{k,i}^{\mathrm{proc}}(t)$ is incurred. Therefore, the per-hop delay of task $k$ on hop $(i,j)$ can be written as
\begin{equation}
\tau_{k,i,j}(t)
=
\tau_{i,j}^{\mathrm{prop}}(t)
+
\tau_{k,i,j}^{\mathrm{tx}}(t)
+
\tau_{k,i,p}^{\mathrm{q}}(t)
+
\tau_{k,i}^{\mathrm{proc}}(t),
\end{equation}

Let $\mathcal{P}_k$ denote the multi-hop path traversed by task $k$. Then the cumulative end-to-end transmission delay is
\begin{equation}
D_k=\sum_{(i,j)\in\mathcal{P}_k}\tau_{k,i,j}(t),
\end{equation}

\subsection{Problem Formulation}
Let $I_k$ and $\hat{I}_k$ denote the source image and the reconstructed image of semantic task $k$, respectively. To characterize end-to-end semantic delivery performance, we define a normalized semantic quality score
\begin{equation}
Q_k^{\mathrm{sem}}=\mathcal{N}\!\big(\Phi(I_k,\hat{I}_k)\big), \qquad 0\le Q_k^{\mathrm{sem}}\le 1,
\end{equation}
where $\Phi(\cdot,\cdot)$ denotes the selected semantic quality evaluator and $\mathcal{N}(\cdot)$ denotes a normalization mapping to $[0,1]$.

Let $\mathcal{K}$ denote the set of semantic transmission tasks. The considered problem aims to jointly optimize routing and semantic processing so as to reduce cumulative transmission delay while improving end-to-end semantic quality. We formulate the objective as
\begin{equation}
\begin{aligned}
\min_{\Pi}\quad
& \frac{1}{|\mathcal{K}|}
\sum_{k\in\mathcal{K}}
\mathbb{E}_{\Pi}
\big[
\lambda_d D_k + \lambda_s(1-Q_k^{\mathrm{sem}})
\big] \\
\textrm{s.t.}\quad
& \mathrm{C1:}\ (i,j)\in\mathcal{E}(t), \qquad \forall (i,j)\in \mathcal{P}_k, \\
& \mathrm{C2:}\ 0 \le q_{i,p}(t) \le q_{\max}, \qquad \forall i,p,t, \\
& \mathrm{C3:}\ m_i \in \mathcal{M}, \quad c_i \in \mathcal{C}, \qquad \forall i, \\
& \mathrm{C4:}\ \mathrm{TTL}_k(t)\ge 0, \qquad \forall k,t, \\
& \mathrm{C5:}\ \lambda_d \ge 0,\ \lambda_s \ge 0,\ \lambda_d+\lambda_s=1.
\end{aligned}
\label{eq:problem_formulation}
\end{equation}
where $\Pi$ denotes the overall forwarding-and-processing policy, $\lambda_d\ge 0$ and $\lambda_s\ge 0$ are weighting coefficients, and $\lambda_d+\lambda_s=1$. $\mathcal{M}$ and $\mathcal{C}$ denote the feasible relay-mode set and semantic-budget set, respectively.

\section{The Proposed GraphJSCR Method}

\subsection{GAT-based State Representation Learning}
As shown in Fig.~\ref{fig:framework}, each packet-holding satellite can only access local observations and one-hop neighboring information, which makes direct hop-by-hop decision making susceptible to local optima. To enhance local perception under partial observability, we construct a relay-aware subgraph centered at the current forwarding node and employ a graph attention network (GAT) to aggregate neighboring information.

Specifically, for the current node $i$, let $\mathcal{N}_i(t)$ denote its one-hop neighboring set at time slot $t$, and let $\mathbf{x}_j(t)$ denote the input feature vector of node $j\in\mathcal{N}_i(t)\cup\{i\}$. The subgraph input jointly includes network states, packet states, and semantic states. The attention coefficient from node $j$ to node $i$ is computed as
\begin{equation}
e_{ij}(t)=
\mathrm{LeakyReLU}\!\left(
\mathbf{a}^{\top}
[\mathbf{W}\mathbf{x}_i(t)\,\|\,\mathbf{W}\mathbf{x}_j(t)]
\right),
\end{equation}
\begin{equation}
\alpha_{ij}(t)=
\frac{\exp(e_{ij}(t))}
{\sum_{u\in\mathcal{N}_i(t)\cup\{i\}}\exp(e_{iu}(t))},
\end{equation}
where $\mathbf{W}$ is the learnable projection matrix, $\mathbf{a}$ is the attention vector, $\|$ denotes concatenation. The hidden representation of node $i$ is then obtained by weighted neighborhood aggregation:
\begin{equation}
\mathbf{h}_i(t)=
\sigma\!\left(
\sum_{j\in\mathcal{N}_i(t)\cup\{i\}}
\alpha_{ij}(t)\mathbf{W}\mathbf{x}_j(t)
\right),
\end{equation}
where $\sigma(\cdot)$ is a nonlinear activation function.

The resulting embedding $\mathbf{h}_i(t)$ provides a compact representation that implicitly captures local topology variation, queue/load propagation trends, and semantic transmission context. This GAT-enhanced representation is then fed into the downstream decision network for hop-by-hop joint optimization.

\subsection{GraphJSCR Decision Network}
To model the hop-by-hop joint decision process under local observations, we formulate multi-hop semantic forwarding in the dynamic LEO network as a partially observable Markov decision process (POMDP).

\subsubsection{State Space}
At time slot $t$, the current packet-holding node $i$ receives a local observation
\begin{equation}
o_i(t)=\big[o_i^{\mathrm{net}}(t),\,o_i^{\mathrm{pkt}}(t),\,o_i^{\mathrm{sem}}(t)\big],
\end{equation}
where $o_i^{\mathrm{net}}(t)$ contains network-related states such as node features, local topology, link status, and queue/load status; $o_i^{\mathrm{pkt}}(t)$ contains packet-related states such as current/destination information and port queue status; and $o_i^{\mathrm{sem}}(t)$ contains semantic-related states such as measured SNR and semantic quality proxy. After one-hop graph aggregation, the input to the policy network is written as
\begin{equation}
s_i(t)=\big[o_i(t),\,\mathbf{h}_i(t)\big],
\end{equation}
Therefore, the decision state jointly reflects local network dynamics, packet context, and semantic transmission conditions.

\subsubsection{Action Space}
Instead of flattening all forwarding choices into a single label, GraphJSCR adopts a factorized joint action design:
\begin{equation}
a_i(t)=\Big(a_i^{\mathrm{hop}}(t),\,a_i^{\mathrm{c}}(t),\,a_i^{\mathrm{relay}}(t)\Big),
\end{equation}
where $a_i^{\mathrm{hop}}(t)$ denotes next-hop selection; $a_i^{\mathrm{c}}(t)$ denotes the semantic channel budget, chosen from $C=\{64,96,128\}$, where each value determines the semantic payload capacity; $a_i^{\mathrm{relay}}(t)$ denotes the relay processing level, where 0 means direct forwarding and 1 means relay processing before next-hop transmission. The action can be interpreted as an instruction tuple that jointly determines where the packet should be forwarded, how much semantic transmission budget should be allocated, and whether relay-side semantic processing should be activated.

\subsubsection{Reward Function}
The reward is designed to jointly encourage efficient forwarding and high semantic fidelity. Let $r_t$ denote the immediate reward at time slot $t$. It is defined as
\begin{equation}
r_t = r_t^{\mathrm{tr}} + \beta_{\mathrm{sem}}\, r_t^{\mathrm{sem}},
\end{equation}
where $r_t^{\mathrm{tr}}$ is the transmission-related reward and $r_t^{\mathrm{sem}}$ is the semantic-quality reward, $\beta_{\mathrm{sem}}$ is a weighting coefficient.

The transmission-related reward is further written as
\begin{equation}
r_t^{\mathrm{tr}} = r_t^{\mathrm{prog}} + r_t^{\mathrm{succ}} - r_t^{\mathrm{fail}},
\end{equation}
where $r_t^{\mathrm{prog}}$ is a shaping term for forwarding progress, $
r_t^{\mathrm{prog}}
=
\omega_h \Delta_t^{\mathrm{hop}}
-\omega_d p_t^{\mathrm{delay}}
-\omega_q p_t^{\mathrm{queue}}
-\omega_l p_t^{\mathrm{loop}},$where $\Delta_t^{\mathrm{hop}}$ denotes the reduction of the residual distance to the destination after taking the current action, and $p_t^{\mathrm{delay}}$, $p_t^{\mathrm{queue}}$, and $p_t^{\mathrm{loop}}$ denote the penalties associated with delay growth, queue congestion, and loop risk, respectively.
$r_t^{\mathrm{succ}}$ is the terminal reward for successful delivery, and $r_t^{\mathrm{fail}}$ is the penalty for failed forwarding events such as TTL expiration, queue overflow, or unavailable outgoing links.
$r_t^{\mathrm{sem}}$ is activated when a semantic session is successfully completed.

\begin{algorithm}[t]
\caption{Training Procedure of the Proposed GraphJSCR}
\label{alg:jrsccppo}
\begin{algorithmic}[1]
\STATE Initialize actor parameters $\theta$, critic parameters $\phi$, and rollout buffer $\mathcal{D}$
\FOR{each training episode}
    \STATE Reset the dynamic LEO environment and initialize semantic tasks
    \FOR{each environment step}
        \STATE Observe local state $s_t$ of the packet-holding node
        \STATE Construct the relay-aware subgraph and obtain GAT embedding $\mathbf{h}_t$
        \STATE Generate joint action $a_t=(a_t^{\mathrm{hop}},a_t^{\mathrm{c}},a_t^{\mathrm{relay}})$ from policy $\pi_\theta$
        \STATE Execute $a_t$, and obtain reward $r_t$ and next state $s_{t+1}$
        \STATE Store $(s_t,a_t,r_t,s_{t+1})$ and related information in $\mathcal{D}$
        \IF{the rollout horizon is reached or the episode terminates}
            \STATE Compute return and advantage targets from $\mathcal{D}$
            \FOR{each PPO update epoch}
                \STATE Sample mini-batches from $\mathcal{D}$
                \STATE Update actor $\theta$ using the clipped surrogate objective
                \STATE Update critic $\phi$ using the value loss
            \ENDFOR
            \STATE Clear rollout buffer $\mathcal{D}$
        \ENDIF
    \ENDFOR
\ENDFOR
\STATE Output the trained shared GraphJSCR policy
\end{algorithmic}
\end{algorithm}

\subsection{Training Strategy of GraphJSCR}
GraphJSCR is trained end-to-end in the dynamic LEO simulation environment through repeated environment interaction and policy optimization. At each step, the current node observes its local state, performs a forward pass through the actor network, executes the selected joint action, and receives the resulting reward. The collected rollout samples are then used to update the actor and critic.

To stabilize policy learning, we adopt Proximal Policy Optimization (PPO) with the clipped surrogate objective
\begin{equation}
\mathcal{L}_{\mathrm{clip}}
=
\mathbb{E}_t
\left[
\min\!\Big(
r_t(\theta)\hat{A}_t,\,
\mathrm{clip}(r_t(\theta),1-\epsilon,1+\epsilon)\hat{A}_t
\Big)
\right],
\end{equation}
where
\begin{equation}
r_t(\theta)=
\frac{\pi_{\theta}(a_t\mid s_t)}
{\pi_{\theta_{\mathrm{old}}}(a_t\mid s_t)},
\end{equation}
$\hat{A}_t$ is the advantage estimate, and $\epsilon$ is the clipping coefficient. The full training objective further includes a value regression term and an entropy regularization term:
\begin{equation}
\mathcal{L}
=
-\mathcal{L}_{\mathrm{clip}}
+
c_v \mathcal{L}_{\mathrm{value}}
-
c_e \mathcal{H}\big(\pi_{\theta}\big),
\end{equation}
where $\mathcal{L}_{\mathrm{value}}$ is the critic loss, $\mathcal{H}(\pi_{\theta})$ denotes the policy entropy, and $c_v$ and $c_e$ are weighting coefficients.

During training, actor and critic parameters are updated for multiple epochs using minibatch rollout samples. After convergence, the learned GraphJSCR policy can be deployed in a fully distributed manner, where each packet-holding satellite reuses the shared policy to adaptively coordinate next-hop selection, semantic budget allocation, and relay-side processing under dynamic topology and local observations.

\section{Evaluation}

\subsection{Experiment Setup}
All experiments are conducted in ns-3.41, where the simulator is connected to the learning agent through the ns3-ai Gym interface. We consider a pure ISL scenario with a Walker-like constellation consisting of 10 orbital planes and 7 satellites per plane at an orbital altitude of 570~km. The maximum queue length and packet time-to-live are 600 packets and 16 hops, respectively. A random channel perturbation with a standard deviation of 1~dB is introduced. In addition, a slow time-varying jitter with an amplitude of 2.0~dB is imposed, the channel correlation horizon is set to 2.0~s, and the link failure rate is 0.05.

For semantic transmission, the payload chunk size is 1200 bytes and the semantic frame interval is 6s. The default evaluation uses a single-session regime, with each flow generating at most one semantic session per episode. Semantic performance is evaluated on selected images from the DIV2K dataset. We use SSIM and a CLIP-based image-image cosine similarity metric \cite{b15} to measure structural fidelity and semantic consistency, respectively. We also report session drop rate as an auxiliary reliability metric for failures caused by congestion, TTL expiration, or unavailable forwarding.

For policy training, we adopt PPO with a learning rate of $5\times10^{-5}$, a discount factor of 0.99, a rollout horizon of 256, a clip ratio of 0.2, 4 update epochs, and a mini-batch size of 128. The entropy and value loss coefficients are set to 0.05 and 0.5, respectively. The weighting coefficient $\beta_{\mathrm{sem}}$ is 1.

\subsection{Baseline Algorithms}
We consider two groups of baselines, corresponding to backend comparison and routing-oriented comparison.

For backend comparison, we evaluate the proposed \textit{GraphJSCR} against classical semantic transmission baselines, including \textit{JPEG2000+LDPC} and \textit{DeepJSCC}. This group is used to compare complete end-to-end semantic transmission schemes under the same dynamic LEO environment and task setting, rather than to isolate the coding module alone. In addition, to assess the contribution of relay-aware semantic processing, we compare the full \textit{GraphJSCR} scheme with a reduced variant using plain \textit{SwinJSCC} without relay strategy.

For routing-oriented comparison, the semantic frontend is fixed to \textit{SwinJSCC}. On this basis, we compare different decision mechanisms, including the proposed \textit{GraphJSCR}, \textit{GraphPR}\cite{b3}, and \textit{DQN-IR}\cite{b16}, in order to isolate the gain brought by joint routing, relay processing, and semantic budget adaptation.

For ablation, we test two reduced GraphJSCR variants—without source-$C$ control and without relay processing—to isolate the gains of semantic budget adaptation and relay-aware processing under heavier load.

\begin{figure}[t]
    \centering
    \includegraphics[width=0.65\columnwidth]{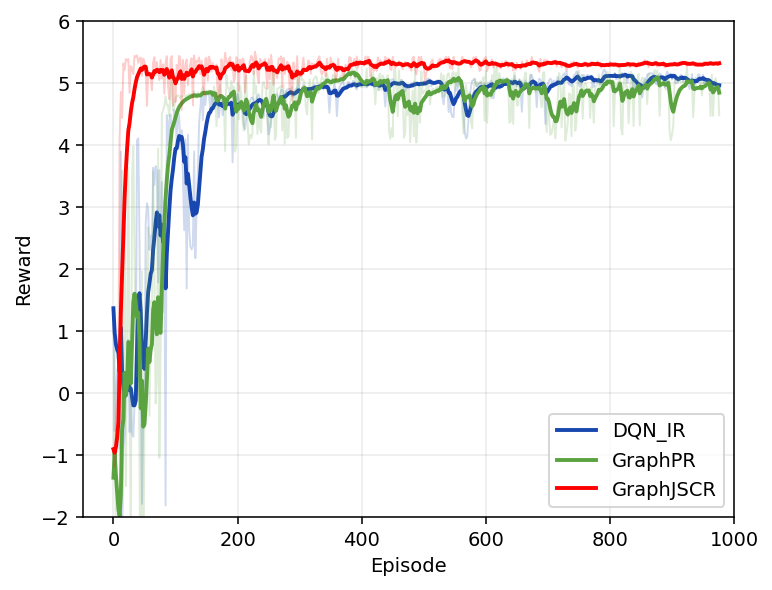}
    \caption{Convergence Analysis.}
    \label{fig:convergence_analysis}
\end{figure}

\subsection{Results and Analysis}
\textit{Convergence Analysis:} Fig.~\ref{fig:convergence_analysis} compares the training convergence behaviors of DQN-IR, GraphPR, and  GraphJSCR. It can be observed that GraphJSCR converges significantly faster to a higher reward level and exhibits smaller oscillations than DQN-IR and GraphPR during training. This improvement stems from our proposed joint optimization, which helps the policy better balance transmission efficiency and semantic fidelity in dynamic multi-hop LEO networks.

\begin{figure}[t]
\centering
\begin{minipage}[t]{0.49\columnwidth}
    \centering
    \includegraphics[width=\linewidth]{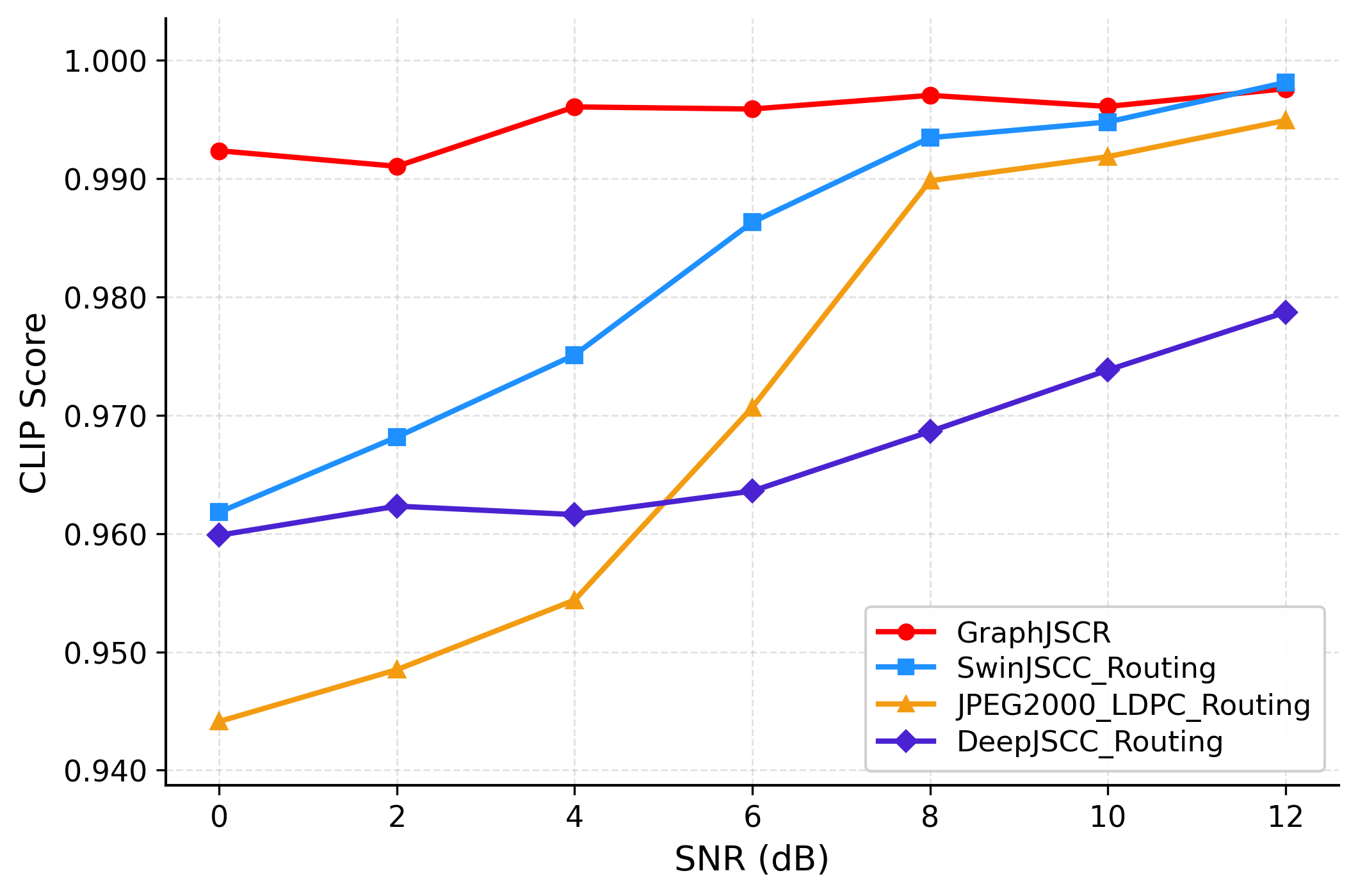}
    \vspace{0.5mm}
    \small (a)
\end{minipage}
\hfill
\begin{minipage}[t]{0.49\columnwidth}
    \centering
    \includegraphics[width=\linewidth]{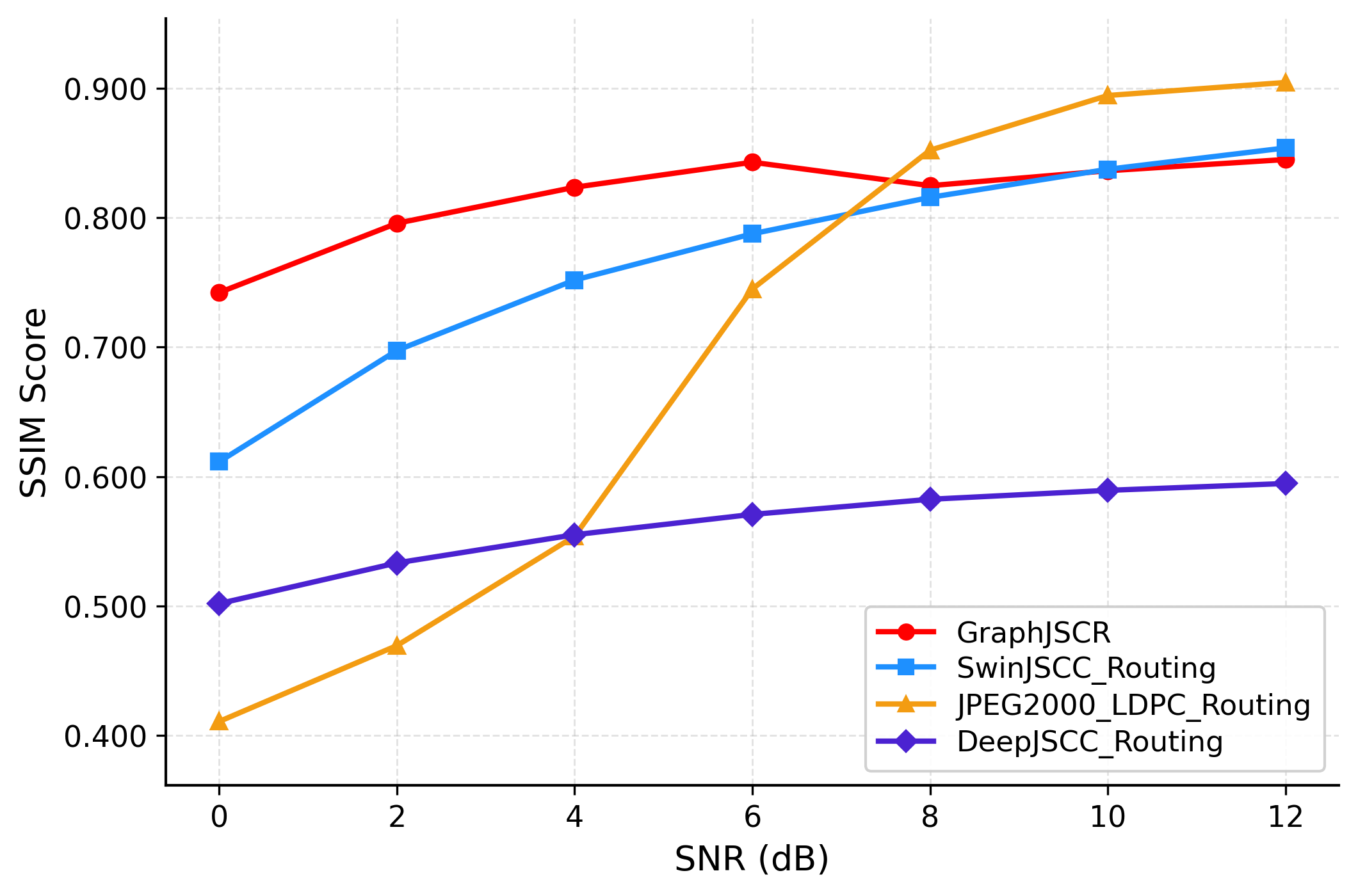}
    \vspace{0.5mm}
    \small (b)
\end{minipage}
\caption{Semantic quality comparison of different schemes under varying SNR conditions. (a) CLIP score. (b) SSIM score.}
\label{fig:snr_quality}
\end{figure}

\textit{Semantic Quality under Different SNRs:} Fig.~\ref{fig:snr_quality}(a) and Fig.~\ref{fig:snr_quality}(b) compare the semantic consistency and structural fidelity of different schemes under varying SNRs. GraphJSCR achieves the best or near-best CLIP score across the whole SNR range, with a particularly clear advantage in the low- and medium-SNR regimes. In terms of SSIM, GraphJSCR also shows superior performance at low SNRs and significantly outperforms other methods in this regime. Although JPEG2000+LDPC performs best at high SNRs, GraphJSCR still maintains competitive quality while avoiding unnecessary semantic payload overhead. This indicates that GraphJSCR prioritizes semantic quality preservation under poor channel conditions, while under favorable channels it tends to reduce transmission load without sacrificing reconstruction quality. Considering both SSIM and CLIP, GraphJSCR provides a more robust and balanced semantic transmission solution across different channel conditions.

\begin{figure}[t]
\centering
\begin{minipage}[t]{0.49\columnwidth}
    \centering
    \includegraphics[width=\linewidth]{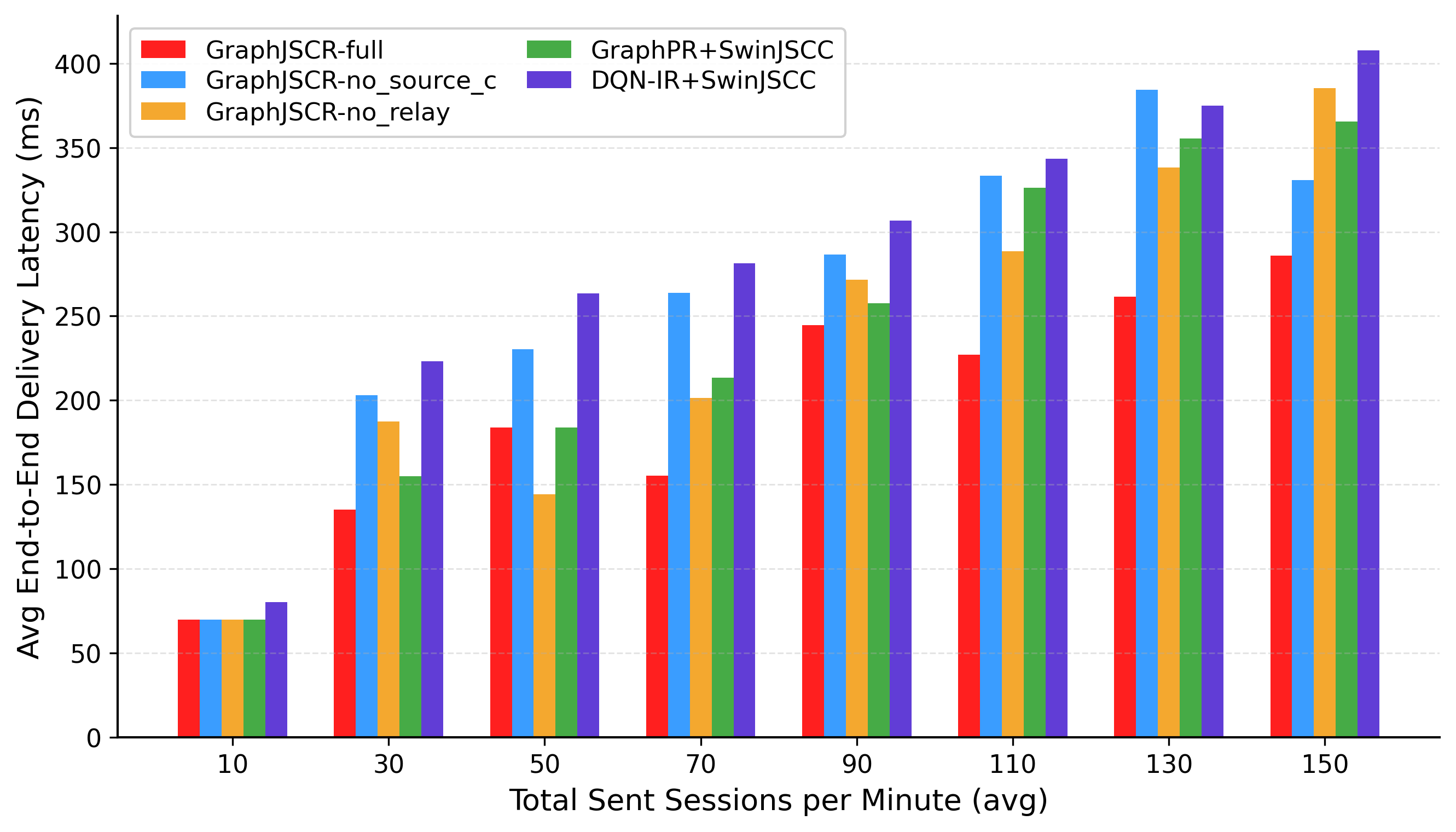}
    \vspace{0.5mm}
    \small (a)
\end{minipage}
\hfill
\begin{minipage}[t]{0.49\columnwidth}
    \centering
    \includegraphics[width=\linewidth]{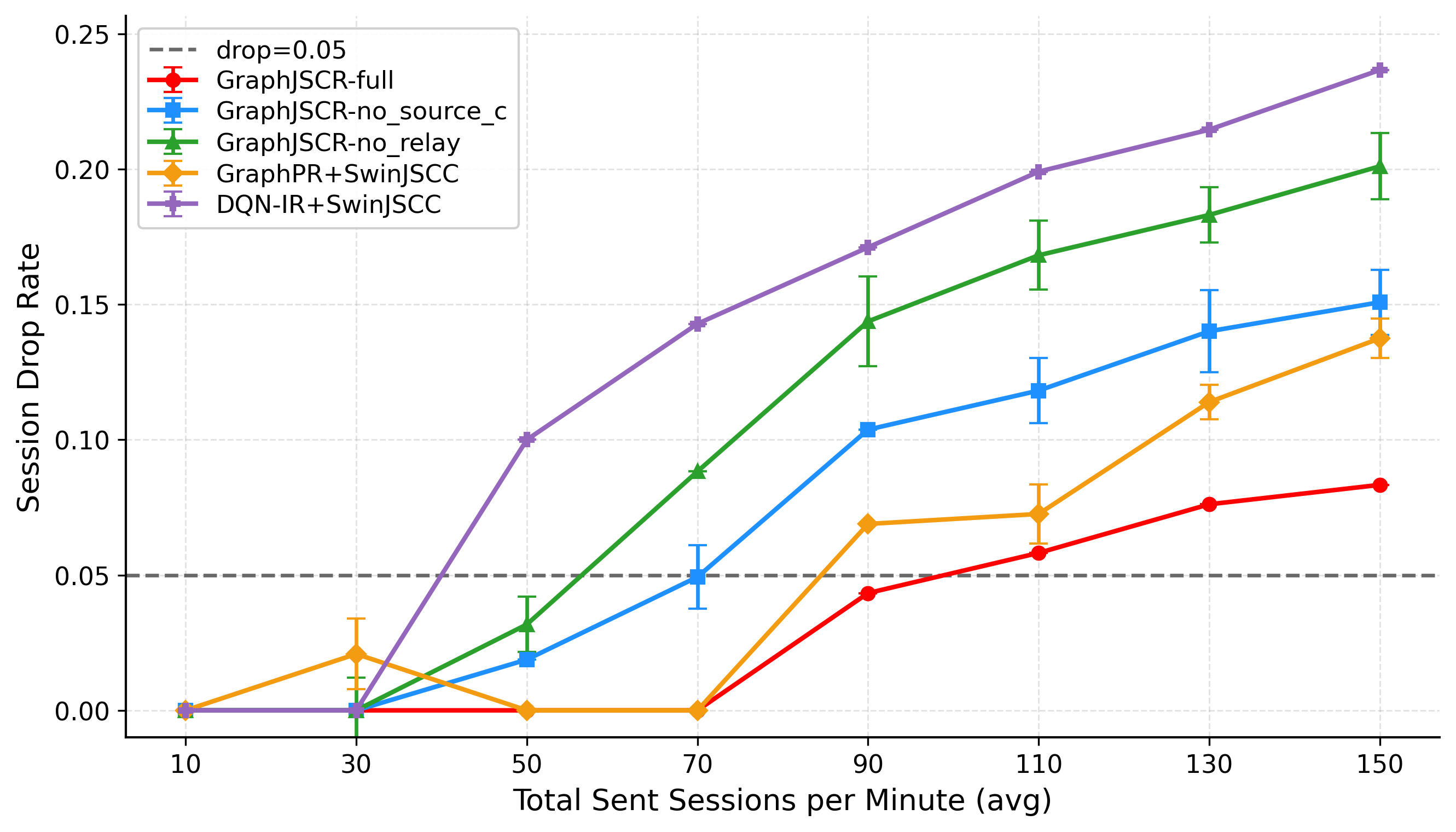}
    \vspace{0.5mm}
    \small (b)
\end{minipage}
\caption{Load sensitivity comparison under different semantic session loads. (a) Average end-to-end delivery latency. (b) Session drop rate.}
\label{fig:load_sensitivity}
\end{figure}

\textit{Load Sensitivity under Increasing Session Load:} Fig.~\ref{fig:load_sensitivity}(a) and Fig.~\ref{fig:load_sensitivity}(b) show the average delivery latency and session drop rate under different semantic session loads, where the load level is controlled by varying the number of concurrent flows while keeping the semantic frame interval and packetization setting fixed. As the offered load increases, all methods suffer from higher queueing delay and reduced delivery reliability. Nevertheless, GraphJSCR-full consistently maintains the lowest overall latency and the most favorable robustness--latency tradeoff in terms of session success. The ablation results further show that removing adaptive source-$C$ control leads to the most noticeable degradation, while removing relay processing mainly hurts robustness under high load. 

\begin{figure}[t]
\centering
\includegraphics[width=\columnwidth]{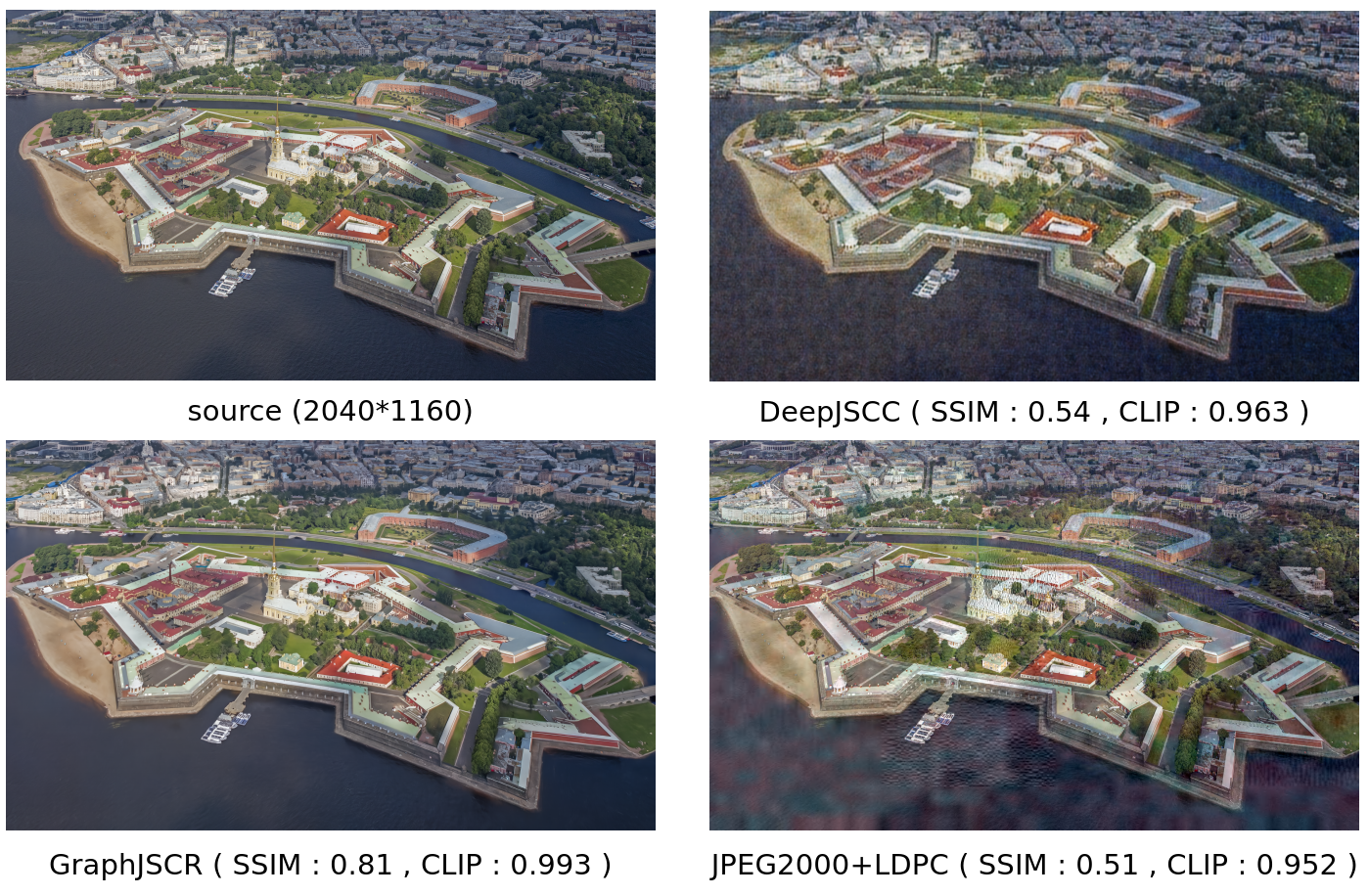}
\caption{Qualitative reconstruction comparison at $\mathrm{SNR}=3$~dB. GraphJSCR employs adaptive source-channel selection over candidate channel budgets $\{64,96,128\}$, and the policy selects $C=128$ at this SNR. DeepJSCC uses latent channel number $c=8$ with an inference patch size of $128\times128$. JPEG2000+LDPC uses compression rate $r=16$ with LDPC parameters $(n=128,d_v=2,d_c=4)$ and a maximum of 40 decoding iterations.}
\label{fig:qualitative}
\end{figure}

\textit{Qualitative Reconstruction Comparison:} Fig.~\ref{fig:qualitative} presents a visual comparison at $\mathrm{SNR}=3$~dB. Under this low-SNR condition, both DeepJSCC and JPEG2000+LDPC suffer from evident quality degradation. DeepJSCC exhibits texture blurring and loss of fine structural details, whereas JPEG2000+LDPC shows artifact contamination and local structural distortion. By contrast, GraphJSCR preserves the overall scene semantics much more faithfully, achieving the highest SSIM and CLIP scores among the compared schemes. Notably, this quality advantage is obtained with only 931 transmitted packets, which is comparable to JPEG2000+LDPC (908 packets) and far smaller than DeepJSCC (4588 packets). This visual evidence demonstrates that the proposed method provides stronger robustness for semantic image transmission in challenging low-SNR multi-hop scenarios.

\section{Conclusion}
This paper proposed GraphJSCR, a graph-based joint routing and semantic coding framework for multi-hop semantic transmission in dynamic LEO satellite networks. By modeling the forwarding process as a partially observable sequential decision problem, the proposed method jointly optimizes next-hop selection, relay processing, and semantic transmission budget based on local topology, queue, packet, and semantic states. Simulation results showed that GraphJSCR achieves faster convergence, better semantic fidelity under varying SNR conditions, and stronger robustness under increasing session loads than benchmark methods.

\bibliographystyle{IEEEtran}
\bibliography{refs}

@article{b1,
  title={Low-Earth orbit satellite constellations for global communication network connectivity},
  author={Lagunas, Eva and Chatzinotas, Symeon and Ottersten, Bj{\"o}rn},
  journal={Nature Reviews Electrical Engineering},
  volume={1},
  number={10},
  pages={656--665},
  year={2024}
}

@article{b2,
  title={Deep joint source-channel coding for wireless image transmission},
  author={Bourtsoulatze, Eirina and Kurka, David Burth and G{\"u}nd{\"u}z, Deniz},
  journal={IEEE Transactions on Cognitive Communications and Networking},
  volume={5},
  number={3},
  pages={567--579},
  year={2019}
}

@article{b3,
  author  = {Y. Ran and Y. Ding and S. Chen and J. Lei and J. Luo},
  title   = {Fully-Distributed Dynamic Packet Routing for {LEO} Satellite Networks: A {GNN}-Enhanced Multi-Agent Reinforcement Learning Approach},
  journal = {IEEE Transactions on Vehicular Technology},
  year    = {2025},
  volume  = {74},
  number  = {3},
  pages   = {5229--5234},
  month   = mar
}

@article{b4,
  author  = {K. Yang and S. Wang and J. Dai and others},
  title   = {{SwinJSCC}: Taming Swin Transformer for Deep Joint Source-Channel Coding},
  journal = {IEEE Transactions on Cognitive Communications and Networking},
  year    = {2024},
  volume  = {11},
  number  = {1},
  pages   = {90--104}
}

@article{b5,
  author  = {C. Bian and Y. Shao and H. Wu and E. Ozfatura and D. G{"u}nd{"u}z},
  title   = {Process-and-Forward: Deep Joint Source-Channel Coding Over Cooperative Relay Networks},
  journal = {IEEE Journal on Selected Areas in Communications},
  year    = {2025},
  volume  = {43},
  number  = {4},
  pages   = {1118--1134},
  month   = apr
}

@article{b7,
  title={Multi-hop Parallel Image Semantic Communication for Distortion Accumulation Mitigation},
  author={Xie, Bingyan and Park, Jihong and Wu, Yongpeng and Zhang, Wenjun and Quek, Tony},
  journal={arXiv preprint arXiv:2510.26844},
  year={2025}
}

@inproceedings{b8,
  title={Semantic forwarding for next generation relay networks},
  author={Arda, Enes and Kutay, Emrecan and Yener, Aylin},
  booktitle={2024 58th Annual Conference on Information Sciences and Systems (CISS)},
  pages={1--6},
  year={2024}
}

@article{b9,
  author  = {W. Lin and Y. Yan and L. Li and others},
  title   = {Semantic-forward relaying: A novel framework toward {6G} cooperative communications},
  journal = {IEEE Communications Letters},
  year    = {2024},
  volume  = {28},
  number  = {3},
  pages   = {518--522}
}

@inproceedings{b10,
  author    = {J. Chen and S. Yang and T. T. Chan and others},
  title     = {{SemFusion}: Multi-Source Semantic Information Fusion and Communication},
  booktitle = {2024 International Wireless Communications and Mobile Computing (IWCMC)},
  year      = {2024},
  pages     = {1740--1745},
  publisher = {IEEE}
}

@inproceedings{b11,
  author    = {B. Guo and Z. Xiong and B. Wang and others},
  title     = {Semantic communication-aware end-to-end routing in large-scale {LEO} satellite networks},
  booktitle = {2024 IEEE International Conference on Metaverse Computing, Networking, and Applications (MetaCom)},
  year      = {2024},
  pages     = {137--142},
  publisher = {IEEE}
}

@article{b13,
  author  = {R. Gao and B. Zhang and Q. Zhang and Z. Yang},
  title   = {Task-Oriented Semantic Delivery in Large-Scale Heterogeneous Satellite Networks: A Local-Topological-Information-Dependable Deep Learning Approach},
  journal = {IEEE Internet of Things Journal},
  year    = {2025},
  volume  = {12},
  number  = {21},
  pages   = {45895--45908},
  month   = nov
}

@article{b14,
  author  = {C. Bian and Y. Shao and D. G{"u}nd{"u}z},
  title   = {A Deep Joint Source-Channel Coding Scheme for Hybrid Mobile Multi-Hop Networks},
  journal = {IEEE Journal on Selected Areas in Communications},
  year    = {2025},
  volume  = {43},
  number  = {7},
  pages   = {2543--2559},
  month   = jul
}

@inproceedings{b15,
  author    = {A. Radford and J. W. Kim and C. Hallacy and A. Ramesh and G. Goh and S. Agarwal and G. Sastry and A. Askell and P. Mishkin and J. Clark and G. Krueger and I. Sutskever},
  title     = {Learning Transferable Visual Models From Natural Language Supervision},
  booktitle = {Proceedings of the 38th International Conference on Machine Learning},
  year      = {2021},
  volume    = {139},
  pages     = {8748--8763},
  month     = jul
}

@inproceedings{b16,
  author    = {P. Zuo and C. Wang and Z. Yao and S. Hou and H. Jiang},
  title     = {An intelligent routing algorithm for {LEO} satellites based on deep reinforcement learning},
  booktitle = {Proc. IEEE 94th Vehicular Technology Conference (VTC2021-Fall)},
  year      = {2021},
  pages     = {1--5}
}

\end{document}